\documentclass[%
 reprint,
superscriptaddress,
 amsmath,amssymb,
 aps, prd
]{revtex4-1}

\usepackage{dcolumn}
\usepackage{bm}
\usepackage[dvipdfmx]{graphicx,color,hyperref}

\def\Tr{{\rm Tr}}

\begin{document}


\title{
Critical Dimension and Negative Specific Heat \\
in One-dimensional Large-$N$ Reduced Models}


\author{Takeshi {\sc Morita}}
\email[]{morita.takeshi(at)shizuoka.ac.jp}
\affiliation{ Department of Physics, Shizuoka University, 836 Ohya, Suruga-ku, Shizuoka 422-8529, Japan }
\affiliation{ 
		Graduate School of Science and Technology, Shizuoka University,
		836 Ohya, Suruga-ku, Shizuoka 422-8529, Japan}
\author{Hiroki  {\sc Yoshida}}
\email[]{yoshida.hiroki.16(at)shizuoka.ac.jp}
\affiliation{ Department of Physics, Shizuoka University, 836 Ohya, Suruga-ku, Shizuoka 422-8529, Japan }
\affiliation{ 
	Graduate School of Science and Technology, Shizuoka University,
	836 Ohya, Suruga-ku, Shizuoka 422-8529, Japan}

\date{\today}

\begin{abstract}
		We  investigate critical phenomena of the Yang-Mills (YM) type one-dimensional matrix model that is a large-$N$ reduction (or dimensional reduction) of the $D+1$ dimensional $U(N)$ pure YM  theory  (bosonic BFSS model).
This model shows a large-$N$ phase transition at finite temperature, which is analogous to the confinement/deconfinement transition of the original YM theory.
We study the matrix model at a three-loop calculation via  the ``principle of minimum sensitivity'' and find that there is a critical dimension $D=35.5$:
At $D \le 35$, the transition is of first order, while it is of second order at $D\ge 36$.
Furthermore, we evaluate several observables in our method, and they nicely reproduce the existing Monte Carlo results.
Through the gauge/gravity correspondence, the transition is expected to be related to  a Gregory-Laflamme transition in gravity, and we argue that the existence of the critical dimension is qualitatively consistent with it.
Besides, in the first order transition case, a stable phase having negative specific heat appears in the microcanonical ensemble, which is similar to Schwarzschild black holes. 
We study some properties of this phase.

\end{abstract}

\maketitle


\section{Introduction}

Critical phenomena in physics sometimes show interesting dependences on the numbers of the spatial dimensions.
One remarkable example is the Gregory-Laflamme (GL) transition in the $D+1$ dimensional gravity with a compact $S^1$ circle \cite{Gregory:1994bj}. (See a review \cite{Kol:2004ww}.)
By changing the size of the $S^1$ from small to large, the stable configuration for a given energy changes from a  uniform black string (UBS) to a localized black hole (LBH), and this transition is called the GL transition.
A non-uniform black string (NUBS) may appear as an intermediate state in this transition.
Surprisingly, the order of this phase transition does depend on $D$, and it is of first order at $D \le 12$, while is of second order at $D \ge 13$ \cite{Sorkin:2004qq}.  
Hence, $D=12.5$ can be regarded as  a critical dimension of this transition.
Curiously, if we fix the temperature instead of the energy, the critical dimension changes to $D= 11.5$ \cite{Kudoh:2005hf}.
See Table \ref{Tab-results}.

A similar critical dimension  appears in the Rayleigh-Plateau (RP) instabilities in liquid too. 
If we consider a space time ${\mathbb R}^{D-1,1} \times S^1$ and set a liquid winding the $S^1$ with the same configuration as the UBS.
Suppose that the volume of the liquid is fixed and the radius of the $S^1$ is increased.
(Thus, the liquid is stretched along the $S^1$.)
Then, above a critical radius, this configuration becomes unstable due to the RP instability, and it tends to be non-uniform.
The order of this transition depends on $D$ similar to the GL transition, and it turned out that the critical dimension is $D=11.5$  \cite{Cardoso:2006ks, Miyamoto:2008rd}.
The connection between the GL and RP instabilities was also argued in \cite{Cardoso:2006ks}.

\begin{table}
	\begin{center} 
		\begin{tabular}{|c||c|}
			\hline
			& Critical dimension  \\ \hline \hline
			GL (fixed mass) & 12.5  \\
			\hline
			GL (fixed temperature) & 11.5  \\
			\hline
			RP  & 11.5  \\
			\hline \hline
			YM type matrix model (3-loop) & 35.5  \\
			\hline
		\end{tabular}
		\caption{
			The critical dimensions of various models. 
			The systems show the first order phase transitions below the critical dimensions and they become of second order above them.	
			The critical dimension of the YM type matrix model is one of the main results of this article.
		}
		\label{Tab-results}
	\end{center}
\end{table}

According to the gauge/gravity correspondence \cite{Maldacena:1997re, Itzhaki:1998dd}, the GL transition is expected to be qualitatively related to the confinement/deconfinement (CD) transition in the $D+1$ dimensional Yang-Mills (YM) type matrix quantum mechanics, whose action at finite temperature is given by \cite{Aharony:2004ig, Catterall:2010fx, Mandal:2011hb, Mandal:2011ws, Morita:2014ypa, Dias:2017uyv, Catterall:2017lub},
\begin{align}
	\label{BFSS} 
	S
	=
	\int_0^{\beta} \hspace{-2mm} dt  
	\Tr 
	\Biggl\{ 
	\sum_{I=1}^D
	\frac{1}{2} 
	\left( 
	D_t X^I \right)^2
	-
	\sum_{I,J=1}^D \frac{g^2}{4} [X^I,X^J]^2
	\Biggr\}. 
\end{align} 
This model is a large-$N$ reduction (or dimensional reduction) of the $D+1$ dimensional U($N$) pure Yang-Mills (YM) theory to one dimension \cite{Eguchi:1982nm}.
Here $X^I$ ($I=1,\cdots,D$) are the $N \times N$ Hermitian matrices that are the dimensional reductions of the spatial components of the original $D+1$ dimensional gauge fields.
$D_t:= \partial_t -i [A_t,\,]$ is the covariant derivative and $A_t$ is the gauge field.
$g$ is the coupling constant, and we take the 't Hooft limit $N \to \infty$ and $g\to 0$ with a fixed 't Hooft coupling $\lambda := g^2N$.
Note that this model appears as low energy effective theories of D-branes and membranes in string theories in various situations, and is important in its own right \cite{Aharony:2004ig, deWit:1988wri, Banks:1996vh, Berenstein:2002jq, Aharony:2005ew, Hashimoto:2010je, Hashimoto:2011nm}.

This model shows a large-$N$ phase transition \cite{Sundborg:1999ue, Aharony:2003sx}, which is an analog of the CD transition of the original YM theory \cite{Aharony:2004ig, Aharony:2005ew, Kabat:1999hp, AlvarezGaume:2005fv, Hanada:2007wn, Kawahara:2007fn, Azeyanagi:2009zf, Mandal:2009vz, Azuma:2014cfa, Filev:2015hia, Hanada:2016qbz, Bergner:2019rca, Asano:2020yry}.
The order parameter of this transition is the Polyakov loop operators,
\begin{align}
	u_n := \frac{1}{N} \Tr \exp\left( i n \int_0^\beta dt A_t \right), \qquad(n=1,2,\cdots).
	\label{polyakov}
\end{align}
If $\langle u_n \rangle =0$, ($\forall n$), it indicates a confinement, and,  $\langle u_n \rangle \neq 0$, ($\exists n$) shows a deconfinement.

\begin{figure}
	\begin{center} 
				\includegraphics[scale=0.4]{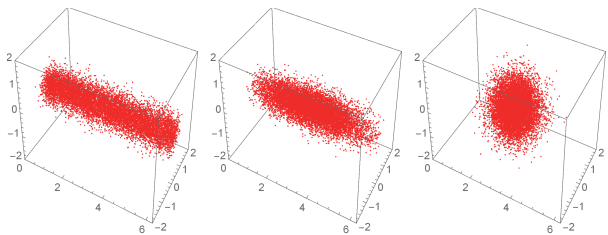}
	
		\caption{Schematic plots of the ``fluids" of the YM matrix model. 
			Their distribution would be uniform, non-uniform or localized along the temporal circle direction.
			These are similar to the black string/black hole systems in gravity.
		}
		\label{fig-schematic}
	\end{center}
\end{figure}

The relation between the CD transition and the GL transition can be intuitively understood as follows.
The diagonal components of $X^I$ can be regarded as the positions of $N$ particles (or D-branes). 
If we take the static diagonal gauge $(A_t)_{ij} = \alpha_i \delta_{ij} $ ($i,j=1,\cdots, N$), 
$\alpha_i$ also describe the positions of the particles.
(Here the configuration space of the gauge field is regarded as a real space.) 
Particularly, the Polyakov loop \eqref{polyakov} is invariant under the shift $\alpha_i=\alpha_i+2\pi/\beta$, and this space is actually an $S^1$ with the period $2\pi/\beta$.
At large-$N$, these particles may behave as a static fluid in the $D+1$ dimension \footnote{Generally, it is non-trivial to compute such a D-brane distribution in matrix models \cite{Hashimoto:2011nm, Azeyanagi:2009zf, Hashimoto:2004fa}.},
and their distribution would be uniform, non-uniform or localized along the $S^1$ as schematically shown in FIG.~\ref{fig-schematic}.
Now the connection to the GL transition in the gravity is clear.
These configurations would correspond to a UBS, NUBS and LBH, respectively.
Note that the temporal component  $\alpha_i$ of the gauge theory corresponds to the spatial $S^1$ direction in the gravity \footnote{
	More precisely,  the temporal direction in the gravity is also periodic in order to make the system to be at a finite temperature. 
	This temporal direction corresponds to the T-dual of one of $X^I$, say $X^1$, in the matrix model. 
	If the temperature in the gravity is sufficiently high, the size of the $S^1$  after the T-dual is large, and we can ignore the periodicity of  $X^1$ in the matrix model \cite{Aharony:2004ig}.
}.
As we have mentioned, the UBS is stable when the size of the $S^1$  is small.
Correspondingly, the uniform distribution in FIG~\ref{fig-schematic} is stable at a small $2\pi/\beta$, which means a low temperature.
We can easily see that  $\langle u_n \rangle =0$ in the uniform distribution, and this is consistent with the confinement at low temperatures.
(The localized distribution is characterized by $u_n \neq 0$ for all $n$ and the non-uniform distribution is characterized by $u_n \neq 0$ for a finite number of $n$'s.
Thus, they are both deconfined.)

Since the critical dimensions appear in the GL and RP transitions, the existence of a critical dimension in the CD transition of
the matrix model is expected.
Indeed, several evidences for this conjecture have been found \cite{Azuma:2014cfa}.
For small $D$, Monte Carlo (MC) simulations show that the order of the CD transition up to $D = 25$ would be of first order  \cite{Azuma:2014cfa, Bergner:2019rca}.
On the other hand, at large-$D$, we can analyze the model analytically through the $1/D$ expansion, and find the second order CD transition \cite{Mandal:2009vz}.
Hence, a critical dimension would exist in the matrix model too.
In this article, we analyze the matrix model by using so called ``principle of minimum sensitivity" \cite{Stevenson:1981vj}, and we will see that the critical dimension is $D=35.5$ at a three-loop calculation.

In addition,  in the first order transition case,  a phase having negative specific heat arises in the microcanonical ensemble \cite{Aharony:2003sx} \footnote{It has been discussed that, generally, some long-range interactions make phases having negative specific heat possible \cite{Thirring1970SystemsWN, LEVIN20141, Berenstein:2018hpl}. 
	In our case, presumably, the interactions between the diagonal components of the matrices may be regarded as long-range, since no screening effect occurs, and the negative specific heat realizes.}.
Since some black holes such as  Schwarzschild black holes and small black holes in AdS space-time  \cite{Hawking:1982dh} have negative specific heat too, the phases in the matrix model would be important to understand why these black holes have negative specific heat from the viewpoint of gauge theories \cite{Aharony:2003sx}.
Although, it is hard to explore such phases in MC calculations, we can easily access this phase in our method. 
We will derive several quantities in this phase near the critical temperature.

\section{Analysis via the Principle of Minimum Sensitivity}

To investigate the phase structure of the model \eqref{BFSS}, we employ the principle of minimum sensitivity \footnote{
	There are several studies, which apply the principle of minimum sensitivity to YM type matrix models \cite{Nishimura:2001sx, Kawai:2002jk, Aoyama:2010ry, Nishimura:2011xy, Hashimoto:2019wmg}.
}.
Such a study was first done by Kabat and Lifschytz \cite{Kabat:1999hp}, but we use a different analysis in order to explore the details of the phase transition.

We deform the model \eqref{BFSS} as
\begin{align}
	\label{modifiedBFSS} 
	S
	=& S_0 + \kappa S_\text{int},  \\
	S_0=&	\int_0^{\beta} \hspace{-2mm} dt  
	\Tr 
	\Biggl\{ 
	\sum_{I=1}^D
	\frac{1}{2} 
	\left( 
	D_t X^I \right)^2
	+\frac{M^2}{2} \left(X^I \right)^2	\Biggr\},
	\nonumber \\
	S_\text{int}=&
	\int_0^{\beta} \hspace{-2mm} dt  
	\Tr 
	\Biggl\{\sum_{I=1}^D -\frac{M^2}{2} \left(X^I \right)^2
	-	\sum_{I,J=1}^D \frac{g^2}{4} [X^I,X^J]^2
	\Biggr\}. \nonumber
\end{align} 
Here we have introduced the deformation parameter $\kappa$ and $M$.
If we take $\kappa=1$, the $M$ dependent terms are canceled, and this model goes back to the original model \eqref{BFSS}.

We integrate out $X^I$ through the perturbative calculations with respect to $\kappa$, and derive the effective action of the Polyakov loop $\{ u_n \}$.
The relevant terms at low temperatures, where all $u_n$ are small \cite{Aharony:2005ew, Aharony:2003sx}, are given by
\begin{align}
	\label{effective-action}
	S_{\text{eff}}(\{ u_n \}, &M)= N^2 \Bigl(\beta f_0 +f_1 |u_1|^2+f_2|u_1|^4 
	\nonumber \\ &
	+ f_3 |u_2|^2 + f_4 (u_2 u_{-1}^2+u_{-2}u_1^2)+ \cdots \Bigr) .
\end{align}
Here $f_0$ is a function of $M$ while $f_i$ ($i=1,2,3,4$) are functions of $M$ and $T:=1/\beta$ \footnote{$f_0$ does not depend on temperature. This is a consequence of the large-$N$ volume independence \cite{Eguchi:1982nm, Gocksch:1982en}. In the confinement phase, all $u_n=0$, and the free energy is given by $N^2 f_0$. The large-$N$ volume independence ensures that the free energy is independent of temperature.}.
The derivations and explicit expressions of $f_0$ and $f_i$ at three-loop order are shown in \eqref{f0} - \eqref{f4} in Appendix \eqref{app-analysis}.
(If we are interested in the two-loop results \footnote{We need at least two-loop to apply the principle of minimum sensitivity.}, we simply remove the terms proportional to $\kappa^2$ in these equations.)

At this stage, we take $\kappa=1$.
Although the initial model \eqref{modifiedBFSS} at $\kappa=1$ is independent of the deformation parameter $M$, the obtained effective action does depend on $M$.
Here, we fix $M$ so that the $M$ dependence of the effective action becomes a minimum.
This prescription is so called ``the principle of minimum sensitivity" \cite{Stevenson:1981vj}. 
Although the validity of such a prescription is generally not ensured, it works very well for many examples.
We will compare our results with the existing studies in order to test our analysis.

\subsection{Low Temperature and Confinement}

To explore the phase structure, we start from considering the low temperature regime.
At low temperatures, we observe $f_1, f_2, f_3 >0$ from \eqref{f1} $\sim$ \eqref{f3}.
Then, the stable configuration in the effective action \eqref{effective-action} is given by $u_1=u_2=0$, and it is in the confinement phase.
Thus, we can approximate $ S_{\text{eff}}= N^2 \beta f_0$, and $M$ at low temperatures is fixed so that the $M$ dependence of $f_0$ is minimized, hence
\begin{align}
	|  \partial_M	f_0 (M=M_0) |  =
	\min |  \partial_M	f_0 (M) |  ,
	\label{eq-M0}
\end{align}	
where $M_0$ denotes the value of $M$ that minimizes $|  \partial_M	f_0|$.
In the two-loop effective action, $f_0 $ has a single extremum $\partial_M f_0=0 $  via \eqref{f0}, and it gives $M_0$ as
\begin{align}
	M_0=\lambda^{1/3}(D-1)^{1/3}, \qquad(\text{two-loop}).
	\label{M0-2-loop}
\end{align}
In the three-loop effective action, $f_0 $ does not have any extremum.
However, it has an  inflection point $\partial^2_M f_0=0 $, which minimizes \eqref{eq-M0}, and we obtain 
\begin{align}
	M_0= \frac{15^{1/3} \lambda^{1/3} }{2}(D-3/4)^{1/3}, \qquad(\text{three-loop}).
	\label{M0-3-loop}
\end{align}

\begin{figure}
	\begin{center} 
				\includegraphics[scale=0.335]{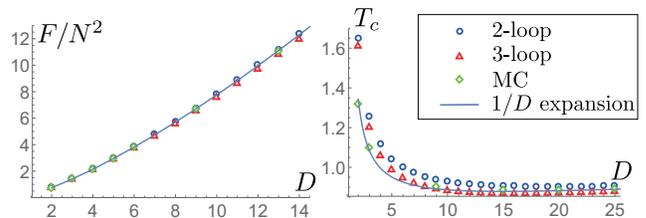}
		\caption{ Free energy $F/N^2$ in the confinement phase (the left panel) and critical temperature $T_c$ (the right panel).
			We have used the unit $\lambda=1$.
			The MC results are from \cite{Azuma:2014cfa, Bergner:2019rca}.
			The $1/D$ expansion results are from (4.27) and (4.30) in \cite{Mandal:2009vz}.
			In the MC results, we plot the transition temperature $T_0$ defined in FIG.\ref{fig-free}, which should be slightly below $T_c$.
			We see good agreement in both of the plots.
		}
		\label{fig-ET}
	\end{center}
\end{figure}

In order to test whether these results are reliable, we evaluate the free energy $F:=S_{\text{eff}}/\beta=N^2 f_0(M_0)$ and compare them with the MC results at low temperatures\footnote{Because of the large-$N$ volume independence \cite{Eguchi:1982nm, Gocksch:1982en}, the temperature dependences of the observables in the confinement phase is very small.
Hence we omit to show temperatures of the MC results in the confinement phase in this article.
}.
By using \eqref{f0}, we obtain $F$ in the confinement phase as
\begin{align}
	F/N^2=
 \left\{
\begin{array}{ll}
	\frac{3}{8} D \lambda^{1/3}(D-1)^{1/3},&\text{(two-loop)}\\
	D \lambda^{1/3}
	\frac{  (1412 D-1187)}{160  ( 30(4 D-3))^{2/3}},&\text{(three-loop)}
	\label{F-2-3-loop}
\end{array}
\right.
.
\end{align}
These results are shown in Fig.~\ref{fig-ET} and Table \ref{Tab-f0-summary}, and both the two- and three-loop analyses show good agreement \footnote{In the $1/D$ expansion \cite{Mandal:2009vz}, it has been shown that the scalar fields $X^I$ acquire a mass dynamically. We guess that the reason for the quantitative success in reproducing the MC results through our analysis is that the mass deformation \eqref{modifiedBFSS} may appropriately capture this dynamical mass.}. 

Furthermore, in Appendix \ref{App-large-D}, we also compare our results \eqref{F-2-3-loop} at large-$D$ with the $1/D$ expansion \cite{Mandal:2009vz}, which would provide reliable results there, and again find good agreement.
Thus, we expect that our analysis via the principle of minimum sensitivity appropriately works in our model \eqref{BFSS}.

\begin{table}
	\begin{center} 
		\begin{tabular}{|c|c|c|c|c|}
			\hline
			$D$	& Two-loop & Three-loop & $1/D$ expansion & MC ($T=0.50$) \\ \hline \hline
			2	& 0.75 & 0.72 & 0.76 &  0.70 ($N=60$) \\ \hline 		
			3	& 1.42 & 1.37 & 1.41 &  1.42 ($N=32$) \\ \hline 				
			4	& 2.16 & 2.09 & 2.15 &  2.11 ($N=32$) \\ \hline 				
			5	& 2.98 & 2.88 & 2.95 &  2.93 ($N=24$) \\ \hline 				
			6	& 3.85 & 3.71 & 3.82 &  3.81 ($N=32$) \\ \hline 				
			9	& 6.75 & 6.52 & 6.71 &  6.66 ($N=32$) \\ \hline 				
			13	& 11.2 & 10.8 & 11.1 &  11.0 ($N=32$) \\ \hline 				
		\end{tabular}
		\caption{
			Free energy $F/N^2$ in the confinement phase.
			We have used the unit $\lambda=1$.
			The two-loop and three-loop results are from \eqref{F-2-3-loop}.
			The $1/D$ expansion results are from \eqref{F-1/D}.
			The MC results are from the unpublished data in \cite{Azuma:2014cfa}.
		}
		\label{Tab-f0-summary}
	\end{center}
\end{table} 

\subsection{Confinement/Deconfinement Transition}

As temperature increases, $f_1(M_0,T)$ becomes negative, and $u_1$ and $u_2$ may obtain non-zero vevs, indicating a deconfinement. 
This is the CD transition in our model.
Near the critical temperature, $u_1$ and $u_2$ would be small and we can perturbatively treat them in the effective action \eqref{effective-action}.
Correspondingly, $M$ can be expanded as
\begin{align}
	\label{M-expansion}
	M=& M_0 +M_1 |u_1|^2+M_2|u_1|^4
	\nonumber \\ &
	+ M_3 |u_2|^2 + M_4 (u_2 u_{-1}^2+u_{-2}u_1^2)+ \cdots  .
\end{align}
Here, in the two-loop theory, $M_0$ is given by \eqref{M0-2-loop} and $M_i$ ($i=1,\cdots,4$) are fixed through the condition  $ \partial_M	S_{\text{eff}}=0 $ in \eqref{effective-action}. 
In the three-loop theory, $M_0$ is given by \eqref{M0-3-loop} and  the condition  $ \partial^2_M	S_{\text{eff}}=0 $ determines $M_i$.
 (See the details in Appendix \ref{App-loop}.)

Then, by substituting \eqref{M-expansion} into the effective action \eqref{effective-action} and using the small $\{ u_n \}$ expansion, we obtain
\begin{align}
	\label{effective-action-un}
	S_{\text{eff}} & (\{ u_n \})= N^2 \Bigl(\beta f_0 +\bar{f}_1 |u_1|^2+\bar{f}_2|u_1|^4 
	\nonumber \\	&
	+ \bar{f}_3 |u_2|^2 + \bar{f}_4 (u_2 u_{-1}^2+u_{-2}u_1^2)+ \cdots \Bigr) .
\end{align}
Here
\begin{align}
	\label{f-bar}
	\bar{f}_i =& f_i + \beta \left( \partial_M f_0 \right) M_i,   \qquad(i=1,3,4), \\
	\bar{f}_2 =& f_2 + \beta  \left( \partial_M f_0 \right) M_2  + \frac{1}{2} \beta  \left( \partial_M^2 f_0 \right) M_1^2+\left( \partial_M f_1 \right) M_1 ,\nonumber 
\end{align}
where $f_i$ are evaluated at $M=M_0$.
Finally, by integrating out $u_2$, we reach a Landau-Ginzburg type effective action for $u_1$,
\begin{align}
	\label{effective-action-u1}
	S_{\text{eff}}( u_1)=& N^2 \left(\beta f_0 +a(T) |u_1|^2+b(T) |u_1|^4 + \cdots \right), \nonumber \\
	& a(T) := \bar{f}_1  , \qquad b(T):= \bar{f}_2 - \frac{\bar{f}_4^2}{\bar{f}_3}.
\end{align}
Although the explicit formulas for $a(T)$ and $b(T)$ are complicated and we omit to show them, it is straightforward to obtain them from \eqref{f0} - \eqref{f4} by using {\it Mathematica}.

Now we can easily see the phase structure \cite{Aharony:2005ew, AlvarezGaume:2005fv, Aharony:2003sx}.
If $a>0$, $u_1=0$ is (meta-)stable and the system may be confined. 
If $a<0$, $u_1=0$ is unstable and $u_1$ has to develop a non-zero vev, and it is deconfinement.
Thus, we can derive the critical temperature $T_c$ by solving $a(T_c)=0$.
Numerical solutions of this equation are shown in FIG.~\ref{fig-ET} and Table \ref{Tab-Tc-summary}.

\begin{table}
	\begin{center} 
		\begin{tabular}{|c|c|c|c|c|}
			\hline
			$D$	& Two-loop & Three-loop & $1/D$ expansion & MC  \\ \hline \hline
			2	& 1.65 & 1.61 & 1.34 &  1.32  \\ \hline 		
			3	& 1.26 & 1.20 & 1.08 &  1.10  \\ \hline 				
			9	& 0.938 & 0.889 & 0.892 &  0.901 \\ \hline 				
			15	 & 0.906 & 0.867 & 0.879 &  0.884  \\ \hline 				
			20	 & 0.903 & 0.869 & 0.882 &  0.884  \\ \hline 				
			25 	& 0.906 & 0.877 & 0.889 &  0.89  \\ \hline 				
		\end{tabular}
		\caption{
			Critical temperature $T_c$.
			We have used the unit $\lambda=1$.
			The $1/D$ expansion results are from \eqref{Tc-1/D}.
			The MC results are from \cite{Azuma:2014cfa} ($D \le 20$) and \cite{Bergner:2019rca} ($D=25$).
			In the MC results, we show the transition temperature $T_0$, which should be slightly below $T_c$.
		}
		\label{Tab-Tc-summary}
	\end{center}
\end{table} 

\begin{figure}
	\begin{center} 
		\includegraphics[scale=0.8]{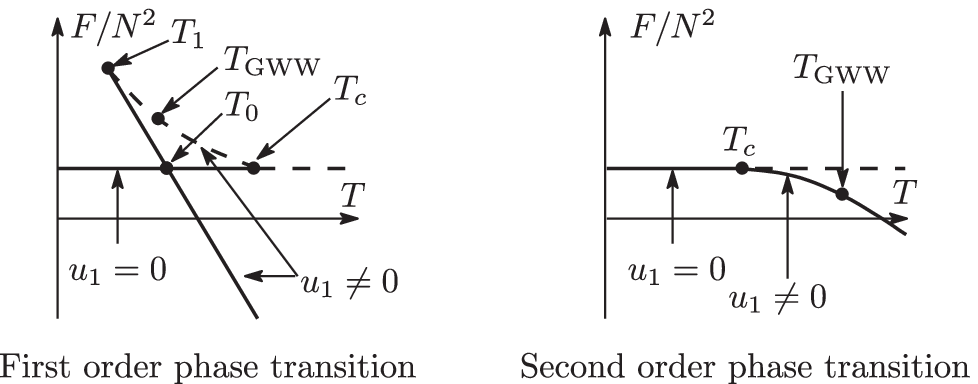}		
		\caption{Schematic plots of  free energy vs. temperature.
	The rigid lines depict stable and meta-stable phases, while the dashed lines depict unstable phases.
	The $u_1=0$ phase lines are horizontal and do not depend on $T$ due to the large-$N$ volume independence.
	In the first order phase transition case (the left panel), the unstable phase with $u_1\neq 0$ merges to the $u_1=0$ branch at $T=T_c$.
	Thus, the $u_1 \neq 0$ solution \eqref{u1} near $T_c$ exists in $T\le T_c$.
	The phase transition occurs not at $T_c$ but at $T_0$ shown in the figure.
	In the second order phase transition case (the right panel), the stable $u_1 \neq 0$ solution \eqref{u1} appears in  $T\ge T_c$.
	Therefore, through \eqref{u1}, the signature of $b$ at $T=T_c$ determines the order of the phase transition.
	Note that there is another transition point $T_{\rm GWW}$, at which a third order transition between the non-uniform distribution and the localized one in FIG.~\ref{fig-schematic} occurs \cite{Aharony:2004ig, Aharony:2005ew, AlvarezGaume:2005fv, Mandal:2009vz}.
	This transition is so called the Gross-Witten-Wadia (GWW) transition \cite{Gross:1980he, Wadia:2012fr}, and is important in the context of the resolution of the naked singularities in the gravity \cite{AlvarezGaume:2005fv, Mandal:2013id}.
}
		\label{fig-free}
	\end{center}
\end{figure}

In order to determine the order of the transition, we expand $a(T)=-c(T-T_c)+\cdots$ ($c:=\partial a/ \partial T>0$) near $T=T_c$, and obtain the classical solution of $u_1$ in \eqref{effective-action-u1} 
as
\begin{align}
	u_1 = \sqrt{-\frac{a}{2b}}  \simeq \sqrt{\frac{c(T-T_c)}{2b}}. 
	\label{u1}
\end{align}
Therefore, if $b(T_c)$ is positive, it indicates a non-trivial solution in $T\ge T_c$, which implies a continuous second order phase transition.
If $b(T_c)$ is negative, an unstable solution exists in $T\le T_c$, and a first order phase transition occurs at a temperature, which is slightly below $T_c$. 
We define this transition temperature as $T_0$.
See FIG.~\ref{fig-free}.

\subsection{Critical Dimension}
\label{sec-critical-dimension}

\begin{figure}
	\begin{center} 
				\includegraphics[scale=0.335]{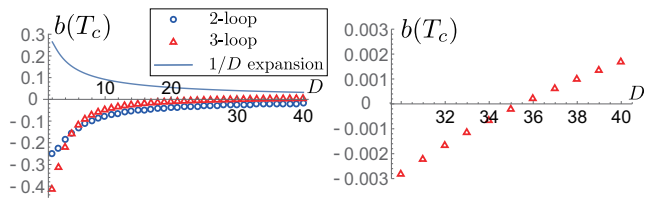}
		\caption{ The value of $b$ at the critical temperature. The negative and positive $b$ indicate the first and second order phase transitions, respectively. 
			At three-loop, $b$ becomes positive at $D=36$, and hence the critical dimension is 35.5. (See the right panel.)
			The $1/D$ result \cite{Mandal:2009vz} is from \eqref{b-1/D}. 
			The MC simulations show first order transitions at least up to $D=25$ \cite{Azuma:2014cfa, Bergner:2019rca}, and they are consistent with our result.
		}
		\label{fig-b}
	\end{center}
\end{figure}

We plot $b(T_c)$ with respect to $D$ in FIG.~\ref{fig-b}.
At two-loop order, $b$ is always negative and it predicts the first order phase transition.
At three-loop order, $b$ becomes positive at $D=36$, and the transition changes to second order.
Thus, the critical dimension of the model \eqref{BFSS} is $D=35.5$ at three-loop.

\begin{figure}
	\begin{center}
		\includegraphics[scale=0.335]{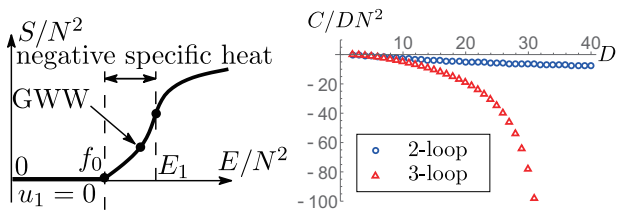}
		\caption{(Left panel) A schematic plot of  entropy vs. energy in the microcanonical ensemble corresponding to the first order transition case in FIG.~\ref{fig-free}.
			The phase having negative specific heat characterized by $\partial^2 S/ \partial E^2 >0$ appears between $E=N^2f_0$ and $E_1$, where 
			$E_1$ is energy at $T=T_1$ in FIG.~\ref{fig-free}.
			The location of the GWW point would depend on $D$ and it may appear even above $E_1$.
			(Right panel)
			Negative specific heat $C$ at $T_c$ in the $u_1\neq 0$ phase.
			In the three-loop case, it diverges at the critical dimension, and the negative specific heat phase disappears in $D \ge 36 $.
		}
		\label{fig-C}
	\end{center}
\end{figure}

\begin{table}
	\begin{center} 
		\begin{tabular}{|c|c|c|c|}
			\hline
			$D$	& Two-loop & Three-loop &  $1/D$ expansion  \\ \hline \hline
			2	& --0.4 & --0.25 & --0.45  \\ \hline 		
			3	& --0.7 & --0.52 & 0.13 \\ \hline 				
			9	& --2.6 & --4.2 & 2.4 \\ \hline 				
			25	& --6.1 & --35.1 & 4.4 \\ \hline 				
			35	& --7.4 & --930 & 5.0 \\ \hline 				
			36	& --7.5 & 1069 & 5.1 \\ \hline 				
			40	& --7.9 & 122 & 5.2 \\ \hline 				
			100	& --11.7 & 16.2 & 6.7 \\ \hline 				
			500	& --17.9 & 6.42 & 9.3 \\ \hline 				
		\end{tabular}
		\caption{
			Specific heat $C/DN^2$ of the $u_1 \neq 0$ phase at $T=T_c$. 
			It is positive in the second order phase transition case, while it is negative in the first order transition case.
			Also, it tends to diverge at the critical dimension $D=35.5$ in the three-loop case.
			The $1/D$ expansion result is from \eqref{C-1/D}.
			The negative specific heat at $D=2$ in the $1/D$ expansion is an error due to the failure of the expansion.
			At $D=2$, the sub-leading term in $b$ \eqref{b-1/D} becomes larger than the leading term and it makes $1/b$ negative after the expansion although $b$ is positive.
			Note that no one has succeeded in the computation of the specific heat at $T=T_c$ via MC.
		}
		\label{Tab-C}
	\end{center}
\end{table} 

\subsection{Phase with Negative Specific Heat}
In the first order transition case, the unstable branch with $u_1 \neq 0$ exists between $T_1$ and $T_c$ as shown in FIG.~\ref{fig-free}.
Remarkably, it becomes stable in the microcanonical ensemble \cite{Aharony:2003sx},  and has negative specific heat akin to a Schwarzschild black hole.
See the schematic plot in FIG.~\ref{fig-C}, where $\partial^2 S/ \partial E^2 >0$ indicates the negative specific heat.
(Here $S$ and $E$ are entropy and energy in the microcanonical ensemble.)
We can read off the specific heat $C:= \partial_T E$ near $T=T_c$ ($T\le T_c$) in this phase via \eqref{effective-action-u1} and \eqref{u1} through the ordinary thermodynamical relations,
\begin{align}
	F&=N^2f_0 - \frac{ N^2 c^2T_c}{4b}(T-T_c)^2+ \cdots,  \nonumber \\
	S&=\frac{E-N^2 f_0}{T_c}- \frac{b}{ c^2 T_0^4} (E-N^2 f_0)^2+ \cdots, \nonumber \\
	C&= \partial_T E=  \frac{ N^2 T^2_c c^2}{2b} + \cdots.
	\label{thermo-quantities}
\end{align}
Here $C$ is negative because $b<0$.
We plot $C$ in FIG.~\ref{fig-C} (right).
See also Table \ref{Tab-C}.
At three-loop, as $D$ approaches to the critical dimension $D=35.5$, $C$ diverges, since $b \to 0$.

\section{Discussions}
We have shown that the critical dimension of the matrix model \eqref{BFSS} is $D=35.5$ at three-loop.
The existence of a critical dimension has been predicted through the MC \cite{Azuma:2014cfa} and the $1/D$ expansion \cite{Mandal:2009vz}, and our result is consistent with them. 
Besides, the strong similarity between the GL, RP and the CD in the matrix model \eqref{BFSS} are sharpened.
This similarity may arise because the matrix model may describe a kind of fluid as depicted in FIG.~\ref{fig-schematic}.
(The obtained critical dimension is different from the gravity, but it would not be a problem because we cannot expect any quantitative agreement in this correspondence \cite{Aharony:2004ig, Mandal:2011ws}.)

However, our analysis relies on the perturbative calculation and the principle of the minimum sensitivity, and $D=35.5$ is not conclusive.
We need the higher order loop calculations to ensure it.
At large-$D$, these corrections may make our results closer to those of the $1/D$ expansion \cite{Mandal:2009vz}. (See Appendix \ref{App-large-D} for the results at large-$D$ in our analysis.)

Also, there are several varieties of the principle of the minimum sensitivity \cite{Hashimoto:2019wmg}, and we need to check whether our results depend on these schemes.

Another remaining problem is understanding the properties of the first order phase transition at $T_0$ in $D \le 35$.
Above $T_0$, the stable configuration would be a non-uniform distribution or a localized one depending on $D$ \footnote{In the second order phase transition case ($D\ge 36$), the stable configuration just above $T_c$ is the non-uniform distribution \cite{Aharony:2005ew, AlvarezGaume:2005fv, Mandal:2009vz, Aharony:2003sx}.}.
If the stable configuration is a non-uniform distribution, another phase transition to a localized distribution must occur at a higher temperature.
Indeed, these transitions have been found in the GL and RP transitions \cite{Kol:2004ww, Miyamoto:2008rd, Kudoh:2004hs, Figueras:2012xj}.
Besides, they would be important for a deeper understanding of the negative specific heat phase in the microcanonical ensemble.

In order to investigate them, we need to evaluate the effective action at finite $\{u_n\}$, and thus we cannot use the expansion \eqref{M-expansion}.
Besides, we need to calculate higher order couplings of the Polyakov loops such as $|u_1|^6$ in the effective action \eqref{effective-action}.
We leave this problem for future work.

\begin{acknowledgments}
We thank Y.~Asano, T.~Azuma, K.~Hashimoto, G.~Mandal, Y. Matsuo, K.~Sugiyama and H.~Suzuki for valuable discussions and comments.
The work of T.~M. is supported in part by  Grant-in-Aid for Young Scientists B (No. 15K17643) from JSPS.
\end{acknowledgments}

\appendix

\section{The Effective Action \eqref{effective-action} at Three-loop}
\label{App-loop}

\begin{figure}[t]
	\begin{center} 
		\includegraphics[scale=1]{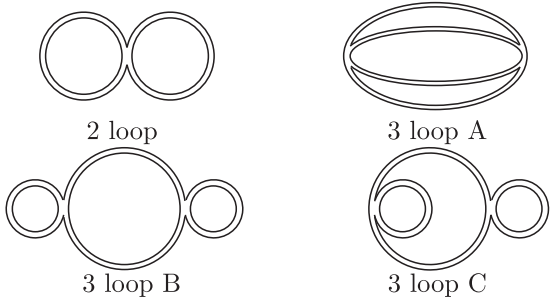}
		\caption{Planar diagrams to compute the effective action.
			At three-loop order, the diagrams A, B and C correspond to $S^A_{\text{3-loop}}$, $S^B_{\text{3-loop}}$ and $S^C_{\text{3-loop}}$ in \eqref{3loop-44}, respectively.
		}
		\label{fig-diagram}
	\end{center}
\end{figure}

In this Appendix, we derive the effective action \eqref{effective-action} at three-loop order.
Starting from the deformed action \eqref{modifiedBFSS}, we compute the effective action of the Polyakov loop  $\{ u_n \}$ by integrating out $X^I$ through the standard perturbative calculation with respect to $\kappa$.
It will lead to the expansion,
\begin{align}
	S_{\text{eff}}(\{ u_n \},M)=  \sum_{m=1}^3 \kappa^{m-1} S_{m\text{-loop}} .
	\label{kappa-expansion}
\end{align}
The analysis mainly follows that of the massive BFSS model \cite{Aharony:2005ew}.
In order to compute this expansion, we use the propagator of $X^I$
in the static diagonal gauge $(A_t)_{ij} = \alpha_i \delta_{ij} $ 
\cite{Mandal:2009vz},
\begin{align} 
	&\langle X_{ij}^I(t) X_{kl}^J(0) \rangle= 
	\delta_{il} \delta_{jk} \delta^{IJ} \frac{1}{2M}	e^{ i    (\alpha_i-\alpha_j)||t||} \nonumber \\
	& \times
	\Biggl[
	e^{ -M ||t||	}
	\sum_{n=0}^{\infty}x^n u^i_n u_{-n}^{j}+
	e^{ 
		M ||t||
	}
	\sum_{n=1}^{\infty}x^n u^{i}_{-n} u_n^{j}
	\Biggr]
	.
	\label{propagator with A}
\end{align} 
Here $x := e^{-\beta M}$ and $||t||$ denotes $||t+n \beta||=t$ for $0\le t < \beta$.
$u_n^i := e^{i \beta n \alpha_i}$,
which satisfies $\sum_{i=1}^{N} u_n^i =N u_n$, where $u_n$ is the $n$-th Polyakov loop defined in \eqref{polyakov}.

Through the one-loop integral, we obtain
\begin{align}
	S_{\text{1-loop}}/N^2 = \frac{D\beta M}{2}
	+ \sum_{n=1}^{\infty}\frac{1-D x^{n}}{n}|u_{n}|^{2}.
	\label{1-loop}
\end{align}
At two-loop, we obtain
\begin{align}
	S_{\text{2-loop}} &=\left\langle\int_0^{\beta} \hspace{-2mm} dt   \Tr \left( -\frac{g^{2}}{4}\left[X^{I},X^{J}\right]^{2}- \frac{M^{2}}{2}\left(X^{I}\right)^2 \right) \right\rangle,
	\label{2-loop}
\end{align}
where
\begin{align}
	&\left\langle\int_0^{\beta} 
	\hspace{-2mm} 
	dt   \Tr \left( -\frac{g^{2}}{4}\left[X^{I},X^{J}\right]^{2} \right) \right\rangle \nonumber \\ &=
	\frac{\beta N^{2}\lambda}{8M^2}D(D-1) 
	+\frac{\beta N^{2}\lambda}{4M^2}D(D-1) \sum_{n=1}^{\infty}(x^{2n}+2x^{n})|u_{n}|^{2} \nonumber \\
	&+\frac{\beta N^{2}\lambda}{8M^2}D(D-1) (x^{2}+2x^{3})(u^{2}_{1}u_{-2}+u^{2}_{-1}u_{2})+\cdots,
	\label{2-loop-4pt}
\end{align}
\begin{align}
	& \left\langle \int_0^{\beta} \hspace{-3mm} dt   \Tr \left(  - \frac{M^{2}}{2}\left(X^{I}\right)^2 \right) \right\rangle
	= - M^2 \frac{\partial}{\partial (M^2)} S_{\text{1-loop}} 
	\nonumber \\ & =
	-\frac{DN^{2} \beta M}{4}-\frac{DN^{2}\beta M}{2}\sum_{n=1}^{\infty}x^{n}|u_{n}|^{2}.
	\label{2-looop-2pt}
\end{align}
Here \eqref{2-loop-4pt} has been computed via the planar diagram depicted in FIG.~\ref{fig-diagram}, and $\cdots$ denotes the irrelevant terms at low temperatures.
On the other hand, \eqref{2-looop-2pt} can be generated from the one-loop result \eqref{1-loop}.

In order to compute the three-loop corrections, we need to evaluate
\begin{widetext}
	\begin{align}
		& -\frac{1}{2} \left\langle \int_0^{\beta} \hspace{-3mm} dt   \Tr \left( -\frac{g^{2}}{4}\left[X^{I},X^{J}\right]^{2} \right) 
		\int_0^{\beta} \hspace{-3mm} dt'   \Tr \left( -\frac{g^{2}}{4}\left[X^{I},X^{J}\right]^{2} \right) 
		\right\rangle_{\text{connected}}
		=S^A_{\text{3-loop}}+S^B_{\text{3-loop}}+S^C_{\text{3-loop}}.
		\label{3loop-44}
	\end{align}
	Here the last three terms are from the three diagrams depicted in FIG.~\ref{fig-diagram}, and we obtain
	\begin{align}\nonumber
		&S^A_{\text{3-loop}}/N^2=-\beta \frac{3\lambda^{2}}{128M^{5}} D(D-1)
		-\frac{3\beta  \lambda^{2}}{32M^{5}}D(D-1)(2\beta Mx^2-x^{3}+x^2+3x)|u_{1}|^{2}\\ &
		-\frac{3 \beta \lambda^{2}}{64M^{5}}D(D-1) (2\beta M x^{2}-2x^{4}+5x^{2})|u_{1}|^{4} 
		-\frac{3 \beta\lambda^{2}}{32M^{5}}D(D-1)\left(4\beta Mx^{4}-x^{6}+x^{4}+3 x^{2}\right)|u_{2}|^{2} \nonumber \\
		&-\frac{3\beta \lambda^{2}}{32M^{5}}D(D-1)\left(4\beta Mx^{3}-x^{5}-x^{4}+3x^{3}+2 x^{2}\right)(u_{2}u_{-1}^{2}+u_{-2}u_{1}^{2})
		+\cdots,
	\end{align}
	\begin{align}
		&S^B_{\text{3-loop}}/N^{2} =-\beta\frac{\lambda^{2}}{32M^{5}} D(D-1)^{2}
		-\frac{\beta\lambda^{2}}{16M^{5}}D(D-1)^{2} \left(\beta M(2x^2+x)+3x^2+3x\right)|u_{1}|^{2} \nonumber \\
		&-\frac{3\beta\lambda^{2}}{16M^{5}}D(D-1)^{2} (\beta M x^{3}+x^{3})|u_{1}|^{4} 
		-\frac{\beta \lambda^{2}}{16M^{5}}D(D-1)^{2}\left(2\beta M\left(2x^{4}+x^{2}\right)+3x^{4}+3x^{2}\right)|u_{2}|^{2} \nonumber \\
		&-\frac{\beta\lambda^{2}}{32 M^5} D(D-1)^{2}\left(2\beta M\left(2x^{4}+3x^{3}+x^{2}\right)+3x^{4}+6x^{3}+3x^{2}\right)(u_{2}u_{-1}^{2}+u_{-2}u_{1}^{2})
		+\cdots,
	\end{align}
	\begin{align}
		&S^C_{\text{3-loop}}/N^2=-\beta\frac{\lambda^{2} }{16M^{5}}D(D-1)^{2} -\frac{\beta \lambda^{2}}{8M^{5}}D(D-1)^{2} \left(\beta Mx(x+1)^{2}+x^{3}+2x^2+3x\right)|u_{1}|^{2}
		\nonumber \\ &
		-\frac{\beta \lambda^{2}}{8M^{5}}D(D-1)^{2} (2\beta Mx^{4}+x^{4}+2x^2)|u_{1}|^{4} 
		-\frac{\beta \lambda^{2}}{8M^{5}}D(D-1)^{2}\left(2\beta M\left(x^{6}+2 x^{4}+x^{2}\right)+x^{6}+2x^{4}+3x^{2}\right)|u_{2}|^{2} \nonumber \\
		&-\frac{\beta \lambda^{2}}{8M^{5}}D(D-1)^{2}\left(\beta M\left(2x^{5}+x^{4}+4x^{3}+x^{2}\right)+x^{5}+x^{4}+3x^{3}+x^{2}\right)(u_{2}u_{-1}^{2}+u_{-2}u_{1}^{2})
		+\cdots.
	\end{align}
	In addition, we need to compute
	\begin{align}
		\label{3loop-42}
		&- \left\langle \int_0^{\beta} \hspace{-3mm} dt   \Tr \left(  - \frac{M^{2}}{2}\left(X^{I}\right)^2 \right)
		\int_0^{\beta} \hspace{-3mm} dt'   \Tr \left( -\frac{g^{2}}{4}\left[X^{I},X^{J}\right]^{2} \right) 
		\right\rangle_{\text{connected}} 
		=-M^2 \frac{\partial}{\partial M^2} \left\langle\int_0^{\beta} \hspace{-3mm} dt   \Tr \left( -\frac{g^{2}}{4}\left[X^{I},X^{J}\right]^{2} \right) \right\rangle \nonumber \\
		=&\frac{\beta N^{2}\lambda}{8M^2}D(D-1) 
		+\frac{\beta N^{2}\lambda}{4M^2}D(D-1) \sum_{n=1}^\infty((n \beta M+1)x^{2n}+(2+n \beta M)x^{n})|u_{n}|^{2} \nonumber \\
		&+\frac{\beta N^{2}\lambda}{8M^2}D(D-1) (( \beta M+1)x^{2}+( 3\beta M+2)x^{3}) 
		(u^{2}_{1}u_{-2}+u^{2}_{-1}u_{2})+\cdots,
	\end{align}
	and
	\begin{align}
		&-\frac{1}{2}\left\langle \int_0^{\beta} \hspace{-3mm} dt   \Tr \left(  - \frac{M^{2}}{2}\left(X^{I}\right)^2 \right)
		\int_0^{\beta} \hspace{-3mm} dt'   \Tr \left(  - \frac{M^{2}}{2}\left(X^{I}\right)^2 \right)
		\right\rangle_{\text{connected}} 
		\nonumber \\
		=& \frac{1}{2} M^4 \frac{\partial^2}{\partial\left( M^2\right)^2} S_{\text{1-loop}} 
		= - \beta \frac{DN^{2}  M}{16}-\frac{DN^{2}}{8}\sum_{n=1}^{\infty}  \left( n\left(\beta M \right)^2+\beta M \right)x^{n}|u_{n}|^{2}.
		\label{3loop-22}
	\end{align}
\end{widetext}
The three-loop correction $S_{3\text{-loop}}$ in \eqref{kappa-expansion} is given as the sum of \eqref{3loop-44}, \eqref{3loop-42} and \eqref{3loop-22}.

By substituting these results to \eqref{kappa-expansion}, we can read off the effective action at three-loop order,
\begin{widetext}
	\begin{align}
		\label{effective-action-app}
		S_{\text{eff}}&(\{ u_n \},M)= N^2 \Bigl(\beta f_0 +f_1 |u_1|^2+f_2|u_1|^4  
		+ f_3 |u_2|^2 + f_4 (u_2 u_{-1}^2+u_{-2}u_1^2)+ \cdots \Bigr) .
	\end{align}
	Here
	\begin{align}
		f_0=&\frac{ DM}{2}
		+ \kappa \left(\frac{\lambda}{8 M^2} D(D-1)-\frac{DM}{4}\right)
		+\kappa^2 \Biggl[-\frac{3 \lambda^2}{128 M^5} D (D-1)-\frac{3 \lambda^2 }{32 M^5}D (D-1)^2
		+ \frac{\lambda}{8 M^2} D(D-1)-\frac{ DM}{16}\Biggr],
		\label{f0}
	\end{align}
	\begin{align}
		f_1=&1-D x+\kappa  \left(\frac{\beta \lambda }{4 M^2} D\left(D-1\right) \left(x^2+2 x\right)-\frac{1}{2} \beta   D M x\right) 
		+\kappa ^2 \Biggl[
		-\frac{3\beta  \lambda^{2}}{32M^{5}}D(D-1)(2\beta Mx^2-x^{3}+x^2+3x)  \nonumber \\ &
		-\frac{\beta\lambda^{2}}{16M^{5}}D(D-1)^{2} \left(\beta M(2x^2+x)+3x^2+3x\right) 
		-\frac{\beta \lambda^{2}}{8M^{5}}D(D-1)^{2} \left(\beta Mx(x+1)^{2}+x^{3}+2x^2+3x\right)
		\nonumber \\ &
		+\frac{\beta \lambda }{4 M^2} D\left(D-1\right) \left(x^2 (\beta  M+1)	+x (\beta  M+2)\right) 
		-\frac{1}{8} D x \left((\beta  M)^2+\beta  M\right)\Biggr],
		\label{f1}
	\end{align}
	\begin{align}
		f_2=&
		\kappa ^2 \Biggl[-\frac{3 \beta \lambda^{2}}{64M^{5}}D(D-1) (2\beta M x^{2}-2x^{4}+5x^{2})
		-\frac{3\beta\lambda^{2}}{16M^{5}}D(D-1)^{2} (\beta M x^{3}+x^{3})
		-\frac{\beta \lambda^{2}}{8M^{5}}D(D-1)^{2} (2\beta Mx^{4}+x^{4}+2x^2) \Biggr],
		\label{f2}
	\end{align}
	\begin{align}
		&f_3=\frac{1}{2} \left(1-D x^2\right)
		+\kappa  \left(\frac{\beta \lambda}{4 M^2}  D\left(D-1\right) \left(x^4+2 x^2\right)-\frac{1}{2} \beta  D M x^2\right)
		+\kappa ^2 \Biggl[
		-\frac{D}{8}  \left( 2\left(\beta M \right)^2+\beta M \right)x^{2} 	\nonumber \\ &
		+\frac{\beta \lambda}{4M^2}D(D-1) \left((2 \beta M+1)x^{4}+(2+2 \beta M)x^{2}\right)
		-\frac{\beta \lambda^{2}}{8M^{5}}D(D-1)^{2}\left(2\beta M\left(x^{6}+2 x^{4}+x^{2}\right)+x^{6}+2x^{4}+3x^{2}\right)
		\nonumber \\ &
		-\frac{\beta \lambda^{2}}{16M^{5}}D(D-1)^{2}\left(2\beta M\left(2x^{4}+x^{2}\right)+3x^{4}+3x^{2}\right)
		-\frac{3 \beta\lambda^{2}}{32M^{5}}D(D-1)\left(4\beta Mx^{4}-x^{6}+x^{4}+3 x^{2}\right)
		\Biggr],
		\label{f3}
	\end{align}
	\begin{align}
		&f_4= \kappa \frac{\beta \lambda}{8M^2}D(D-1) (x^{2}+2x^{3})
		+\kappa ^2 \Biggl[
		\frac{\beta \lambda}{8M^2}D(D-1) (( \beta M+1)x^{2}+( 3\beta M+2)x^{3}) \nonumber \\
		& 
		-\frac{3\beta \lambda^{2}}{32M^{5}}D(D-1)\left(4\beta Mx^{3}-x^{5}-x^{4}+3x^{3}+2 x^{2}\right)-\frac{\beta\lambda^{2}}{32 M^5} D(D-1)^{2}\left(2\beta M\left(2x^{4}+3x^{3}+x^{2}\right)+3x^{4}+6x^{3}+3x^{2}\right) \nonumber \\
		&-\frac{\beta \lambda^{2}}{8M^{5}}D(D-1)^{2}\left(\beta M\left(2x^{5}+x^{4}+4x^{3}+x^{2}\right)+x^{5}+x^{4}+3x^{3}+x^{2}\right)  \Biggr].
		\label{f4}
	\end{align}
\end{widetext}
We have used $x=e^{-\beta M}$.
If we are interested in the two-loop effective action, we should simply ignore $O(\kappa^2)$ terms in this result.

\section{The Details of the Derivation of the Critical Dimension}
\label{app-analysis}
To show the details of the derivations of the critical dimension in Sec.~\ref{sec-critical-dimension}, 
we analyze the effective action \eqref{effective-action-app} and discuss how we determine the phase structure.
We will mainly show the analysis at two-loop, since the three-loop analysis is almost parallel.
(Recall that we remove $O(\kappa^2)$ terms in \eqref{effective-action-app} when we consider the two-loop effective theory.)
We set $\kappa=1$  in \eqref{effective-action-app} hereafter to use the principle of the minimum sensitivity.

First, we consider a low temperature regime.
There, $x=e^{-\beta M}$ would be small, and $f_1$ and $f_3$ would be positive.
Then, to make the effective action \eqref{effective-action-app} small, $\langle u_1 \rangle=\langle u_2 \rangle=0$ would be favored.
Thus, the effective action \eqref{effective-action-app} would become $S_{\text{eff}}=\beta N^2 f_0(M)$.

Here, we need to determine $M$.
As we have discussed in \eqref{eq-M0}, we fix $M$ such that the $M$ dependence of the effective action is minimized.
From \eqref{f0}, we find that at 
\begin{align}
	M_0=\lambda^{1/3}(D-1)^{1/3},
	\label{M0-2-loop-app}
\end{align}
$ \partial_M f_0$ at two-loop becomes 0 and is minimized.
Then we obtain the free energy at low temperatures as 
\begin{align}
	F=S_{\text{eff}}/\beta=N^2f_0(M_0) = \frac{3}{8} N^2 D \lambda^{1/3}(D-1)^{1/3}.
	\label{f0-2-loop}
\end{align}
This result is shown in FIG.~\ref{fig-ET} and Table \ref{Tab-f0-summary}.
We find good agreement with the MC results even at two-loop order.

Next, in order to investigate the phase transition, we compute $M_i$ and $\bar{f}_i$ defined in \eqref{M-expansion} and \eqref{f-bar}.
However, since $ \partial_M f_0=0$ at $M=M_0$, we obtain $\bar{f}_i=f_i$ for  $i=1,3,4$ and we need to evaluate only $M_1$ and $\bar{f}_2$. 
By substituting the expansion \eqref{M-expansion} into the equation $\partial_M S_{\text{eff}}=0$, we find
\begin{align}
	M_1=- \frac{\partial_M f_1}{\beta \partial^2_M f_0} , \qquad \bar{f}_2=-\frac{1}{2}\frac{\left(\partial_M f_1\right)^2}{\beta  \partial^2_M f_0} .
\end{align}
The explicit formulas for these equations are rather messy, and we omit showing them, but one can obtain them easily by using {\it Mathematica}.

Now, we are ready to discuss the critical phenomena.
As we have argued below \eqref{effective-action-u1},
the critical temperature can be found through 
\begin{align}
	0=a(T)=1-De^{-\beta M_0}+\frac{D}{4}\beta M_0 e^{-2\beta M_0}.
	\label{Tc}
\end{align}
This equation can be solved numerically and the result is summarized in FIG.~\ref{fig-ET} and Table \ref{Tab-Tc-summary}.
Again our results seem to be consistent with the MC results. 

Finally, we determine the order of the phase transition.
Through the discussions around \eqref{u1}, it is determined by the signature of $b$ defined in \eqref{effective-action-u1} at the critical temperature.
We numerically see that it is always negative as shown in FIG.~\ref{fig-b} and indicates the first order phase transition for any $D$ at two-loop order.
(As we will shown in \eqref{b-2-loop-large-D}, we can confirm it analytically, if $D$ is large.)
\\

So far, we have shown the two-loop results. Now we will move on to the three-loop case.
The three-loop calculation is almost parallel to the two-loop analysis.
One significant difference is that the minimum of $|\partial_M f_0|$ in \eqref{eq-M0} is not zero.
Hence we need to find the minimum via $\partial_M^2 f_0=0$, and obtain
\begin{align}
	M_0= \frac{15^{1/3} \lambda^{1/3} }{2}(D-3/4)^{1/3}.
	\label{M0-3-loop-app}
\end{align}
The rest of the calculations are straightforward. 
We obtain the free energy in the confinement phase as
\begin{align}
	F=N^2f_0(M_0)=N^2 \lambda^{1/3}
	\frac{ D (1412 D-1187)}{160  ( 30(4 D-3))^{2/3}}.
	\label{f0-3-loop}
\end{align}
This result is shown in FIG.~\ref{fig-ET} and Table \ref{Tab-f0-summary}.

Next, we fix $M_i$ via $\partial_M^2 S_{\text{eff}}=0$ near the critical temperature, and obtain
\begin{align}
	M_i=&- \frac{\partial^2_M f_i}{\beta  \partial^3_M f_0},    \qquad(i=1,3,4), \nonumber \\
	M_2=&- \frac{1}{\beta  \partial^3_M f_0} \nonumber \\
	& \times \left( \partial^2_M f_2-\partial^3_M f_1 \frac{\partial^2_M f_1}{\beta \partial^3_M f_0}   +\frac{\partial^4_M f_0 }{2\beta }  \left(\frac{\partial^2_M f_1}{\partial^3_M f_0}\right)^2 \right) ,
\end{align}
where $f_i$ are evaluated at $M=M_0$.
Then $\bar{f}_i$, $a$ and $b$ are derived through \eqref{f-bar} and \eqref{effective-action-u1}.
Finally, by solving $a(T_c)=0$ and evaluating $b(T_c)$ numerically, we obtain the critical temperatures and the orders of the phase transitions as shown in FIG.~\ref{fig-ET} and \ref{fig-b}.

\section{Other Observables and Specific Heat}

We can also compute other observables via our analysis.
For example, the vevs of the square of the adjoint scalars $X^I$, which have been investigated in the MC studies \cite{Kawahara:2007fn, Azuma:2014cfa}, can be derived as
\begin{align}
	R^2:= & \frac{g^2}{N} \langle \Tr X^I X^I \rangle  \nonumber \\
	= &
	\frac{2 \lambda}{\beta M^2 }\left( - \kappa \partial_\kappa S_{\text{eff}}
	+\lambda \partial_\lambda S_{\text{eff}} 
	\right) |_{\kappa \to 1}
	\label{def-R2},
\end{align} 
where $S_{\text{eff}}$ is the effective action \eqref{kappa-expansion}.
In the confinement phase, it can be calculated as
\begin{align}
	R^2= & \frac{\lambda^{2/3} D}{2 (D-1)^{1/3}}, &(\text{two-loop}), 
	\label{R2-2-loop}
	\\
	R^2= & \frac{\lambda^{2/3} D(148 D -103)}{15 \times 30^{1/3} (4D-3)^{4/3}}, & (\text{three-loop}).
	\label{R2-3-loop}
\end{align}
These quantities agree with the MC studies \cite{Azuma:2014cfa} as shown in Table \ref{Tab-R2}.

\begin{table}
	\begin{center} 
		\begin{tabular}{|c|c|c|c|c|}
			\hline
			$D$	& Two-loop & Three-loop & $1/D$ expansion & MC ($T=0.50$) \\ \hline \hline
			2	& 1.0 & 0.969 & 0.996 &  1.15 ($N=60$) \\ \hline 		
			3	& 1.19 & 1.17 & 1.41 &  1.31 ($N=32$) \\ \hline 				
			4	& 1.39 & 1.37 & 1.42 &  1.45 ($N=32$) \\ \hline 				
			5	& 1.57 & 1.56 & 1.61 &  1.62 ($N=24$) \\ \hline 				
			6	& 1.75 & 1.74 & 1.79 &  1.81 ($N=32$) \\ \hline 				
			9	& 2.25 & 2.24 & 2.28 &  2.29 ($N=32$) \\ \hline 				
			13	& 2.84 & 2.83 & 2.87 &  2.87 ($N=32$) \\ \hline 				
		\end{tabular}
		\caption{
			$R^2$ defined in \eqref{def-R2} in the confinement phase.
			We have used the unit $\lambda=1$.
			The two-loop and three-loop results are from \eqref{R2-2-loop} and \eqref{R2-3-loop}, respectively.
			The $1/D$ expansion results are from \eqref{R-1/D}.
			The MC results are from the unpublished data in \cite{Azuma:2014cfa}.
		}
		\label{Tab-R2}
	\end{center}
\end{table} 

We can also compute specific heat $C$ from the free energy through the ordinary thermodynamical relation, and we obtain
\begin{align}
	C/N^2&= \partial_T E=  O(1/N^2), \quad (T< T_c, ~u_1 =0), \nonumber \\
	&=  \frac{ T^2_c c^2}{2b} \qquad \qquad \qquad ~ (T= T_c,~u_1 \neq 0),
	\label{specific}
\end{align}
where we have used \eqref{thermo-quantities} and 
\begin{align}
	b= \bar{f}_2 - \frac{\bar{f}_4^2}{\bar{f}_3}, \qquad c=\frac{\partial a(T)}{\partial T}.
\end{align}
($a$, $b$, and $\bar{f}_i$ are defined in \eqref{f-bar} and \eqref{effective-action-u1}.)
The specific heat is very small in the confinement phase due to the large-$N$ volume independence, which strongly suppresses temperature dependence of physical quantities \cite{Eguchi:1982nm, Gocksch:1982en}.
It becomes positive in the second order phase transition case, since $b>0$, while it becomes negative in the first order transition case ($b<0$).
Phases with negative specific heat are unphysical in usual thermodynamical systems.
However, in our case, it becomes {\it physical} in the microcanonical ensemble \cite{Aharony:2003sx}.
(See \cite{Thirring1970SystemsWN, LEVIN20141, Berenstein:2018hpl} for some discussions on phases with negative specific heat.)
Also, the specific heat \eqref{specific} at the critical temperature tends to diverge as $D$ approaches to the critical dimension $D=35.5$, where $b$ crosses 0.
(Of course, $D$ is digit, and $b$ cannot be 0.)
The result is summarized in FIG.~\ref{fig-C} and Table \ref{Tab-C}.

\section{Large-$D$ Limit}
\label{App-large-D}

At large-$D$, the $1/D$ expansion \cite{Mandal:2009vz} would be reliable.
Hence, it would be valuable to evaluate our results at large-$D$ and compare them with the $1/D$ expansion \cite{Mandal:2009vz}.

In the large-$D$ expansion, we obtain the following quantities:
\begin{align}
	F/N^2 |_{\beta \to \infty} =&
	D (\lambda D)^{1/3} \nonumber \\
	& \times	 \left(\frac{3}{8}+\frac{1}{D}\left(-\frac{81}{64}+\frac{\sqrt{5}}{2}\right) +O(1/D^2) \right),
	\label{F-1/D}
	\\
	R^2 |_{\beta \to \infty} =&
	(\lambda D)^{2/3} \nonumber \\
	& \times	 \left(\frac{1}{2}+\frac{1}{D}\left(\frac{7 \sqrt{5}}{30}-\frac{9}{32}\right) +O(1/D^2) \right),
	\label{R-1/D}
	\\
	\beta_c=&\frac{\log D}{(\lambda D)^{1/3}}\nonumber \\
	& \times	\left(1+\frac{1}{D}\left(\frac{203}{160}-\frac{\sqrt{5}}{3} \right)+O(1/D^2)\right).
	\label{Tc-1/D}
\end{align}	
These are from (4.27), (4.33) with (4.25) and (4.30) in \cite{Mandal:2009vz}, respectively.
Besides, we evaluate $b(T_c)$ in the effective action \eqref{effective-action-u1}, which fixes the order of the transition, as 
\begin{align}	
	&	b|_{T=T_c}= \frac{\log D}{D} \nonumber \\
	& \times
	\left(\frac{1}{3}+ 
	\frac{1}{D}\left( \frac{1049}{600}- \frac{197\sqrt{5}}{600} - \frac{33}{400} \log D \right)+O(1/D^2)
	\right),
	\label{b-1/D}
\end{align}
where we have used (4.29) and (4.30) in \cite{Mandal:2009vz}.
This is always positive and the $1/D$ expansion predicts the second order phase transition at large-$D$.
Through \eqref{specific}, we obtain the specific heat at $T=T_c$ as
\begin{align}
	&	C/N^2|_{T=T_c}= \frac{3}{2} D \log D \nonumber \\
	&-\frac{3 \log D}{800}
	\left(2098-394 \sqrt{5} -99 \log D \right) +O(1/D).
	\label{C-1/D}
\end{align}
We will compare these quantities with our results at large-$D$.

First, we evaluate our two-loop results at large-$D$.
At two-loop, we can solve \eqref{Tc} at large-$D$ and obtain the critical temperature analytically. 
Then, we obtain
\begin{align}
	F/N^2 |_{\beta \to \infty} =&D(\lambda D)^{1/3}\left(\frac{3}{8}-\frac{1}{8D}+ O(1/D^2)\right), \\
	R^2 |_{\beta \to \infty} =&
	(\lambda D)^{2/3}\left(\frac{1}{2}+\frac{1}{6D}+ O(1/D^2)\right), \\
	\beta_c=&\frac{\log D}{(\lambda D)^{1/3}}\left(1+\frac{1}{12D}+O(1/D^2)\right),  \\
	\quad b|_{T=T_c}=&-\frac{1}{6D}\log D\left(1+O(1/D)\right)  ,
	\label{b-2-loop-large-D}
	\\
	C/N^2|_{T=T_c}=&-3D \log D \left(1+O(1/D)\right).
\end{align}
Thus, $b$ is negative, and it does not agree with the $1/D$ expansion \eqref{b-1/D}.
On the other hand, the leading order terms of $F$, $R^2$ and $\beta_c$ in our results are precisely coincident with those of the $1/D$ expansion, although the $1/D$ corrections differ.
Since the results of the $1/D$ expansion \cite{Mandal:2009vz} would be reliable at large-$D$, these quantities at two-loop order are accidentally very good at large-$D$.

Next, we consider the three-loop results.
Different from the two-loop case, we cannot solve $T_c$ in the three-loop case analytically even at large-$D$.
From \eqref{f0-3-loop} and \eqref{R2-3-loop}, we obtain
\begin{align}
	F/N^2 |_{\beta \to \infty} &=D(\lambda D)^{1/3}\left(\frac{1412}{160 (120)^{2/3}}+ O(1/D)\right), 
	\label{F-3-loop-large-D}
	\\
	R^2 |_{\beta \to \infty} =&
	(\lambda D)^{2/3}\left(\frac{148}{60 (120)^{1/3}}+ O(1/D)\right).
\end{align}
Thus, they do not agree with \eqref{F-1/D} and \eqref{R-1/D} in the $1/D$ expansion.
However, these are numerically not bad. 
For $F$, if we compare the coefficients of the leading terms of \eqref{F-1/D} and \eqref{F-3-loop-large-D}, we obtain $3/8=0.375$ and $1412/160 (120)^{2/3} =0.363...  $ and the error is $3\%$ only.
Similarly, for $R^2$, we have $1/2=0.5$ and $148/60 (120)^{1/3}=0.500092...$, and they are very close.
Hence, we presume that the convergence of the principle of the minimum sensitivity at large-$D$ would be good in our model.

\bibliography{bBFSS}

\end{document}